\documentclass[letterpaper,10pt,conference]{IEEEtran}
\IEEEoverridecommandlockouts
\usepackage{cite}
\usepackage{amsmath,amssymb,amsfonts}
\usepackage{array}
\usepackage{algorithmic}
\usepackage{graphicx}
\usepackage{textcomp}
\usepackage{xcolor}
\usepackage{tabularx}
\usepackage{multicol}  
\usepackage{placeins}
\usepackage{geometry}
\def\BibTeX{{\rm B\kern-.05em{\sc i\kern-.025em b}\kern-.08em
    T\kern-.1667em\lower.7ex\hbox{E}\kern-.125emX}}
\usepackage{microtype}
\usepackage{url} 
\usepackage[breaklinks=true]{hyperref} 
\newcommand{\firstpagemargins}{\geometry{top=1 in, bottom=0.75in, left=0.75in, right=0.75in}} 

\AtBeginDocument{\firstpagemargins}

\begin{document}

\title{\LARGE \bf Hybrid Brain-Machine Interface: Integrating EEG and EMG for Reduced Physical Demand\\
}
\author{Daniel Wang$^1$, Katie Hong$^2$, Zachary Sayyah$^3$, Malcolm Krolick$^2$, \\ 
Emma Steinberg$^4$, Rohan Venkatdas$^5$, Sidharth Pavuluri$^5$, Yipeng Wang$^5$, Zihan Huang$^3$

\thanks{$^{1}$D. Wang (correspondence: dwang118@jhu.edu) is with the Department of Applied Mathematics and Statistics, Johns Hopkins University, Baltimore, MD 21218 USA.}
\thanks{$^{2}$ K. Hong, M. Krolick are with the Department of Computer Science, Johns Hopkins University, Baltimore, MD 21218 USA.}
\thanks{$^{3}$Z. Sayyah, Z. Huang are with the Department of Biomedical Engineering, Johns Hopkins University, Baltimore, MD 21218 USA.}
\thanks{$^{4}$E. Steinberg is with the Department of Electrical Engineering, Johns Hopkins University, Baltimore, MD 21218 USA.}
\thanks{$^{5}$R. Venkatdas, S. Pavuluri, Y. Wang are with the Department of Neuroscience, Johns Hopkins University, Baltimore, MD 21218 USA.}
}
\maketitle

\begin{abstract}

We present a hybrid brain-machine interface (BMI) that integrates steady-state visually evoked potential (SSVEP)-based EEG and facial EMG to improve multimodal control and mitigate fatigue in assistive applications. Traditional BMIs relying solely on EEG or EMG suffer from inherent limitations—EEG-based control requires sustained visual focus, leading to cognitive fatigue, while EMG-based control induces muscular fatigue over time. Our system dynamically alternates between EEG and EMG inputs, using EEG to detect SSVEP signals at 9.75 Hz and 14.25 Hz and EMG from cheek and neck muscles to optimize control based on task demands. In a virtual turtle navigation task, the hybrid system achieved task completion times comparable to an EMG-only approach, while 90\% of users reported reduced or equal physical demand. These findings demonstrate that multimodal BMI systems can enhance usability, reduce strain, and improve long-term adherence in assistive technologies. 

\end{abstract}


\section{Introduction}

Brain-Machine Interfaces (BMIs) have emerged as transformative tools in assistive technology, enabling individuals with severe motor impairments to interact with their environments using neural and physiological signals \cite{Shih, Farina}. However, existing BMI systems predominantly rely on single-modality control, such as electroencephalography (EEG) or electromyography (EMG), each of which presents inherent limitations that reduce long-term usability and adaptability \cite{Nicols, Young}.

EEG-based BMIs, particularly those utilizing steady-state visually evoked potentials (SSVEPs), are commonly used for control interfaces due to their high information transfer rates. However, these systems require sustained visual attention and cognitive engagement, which can lead to cognitive fatigue within 20–30 minutes of continuous use \cite{Trejo, Cao}.

Conversely, EMG-based BMIs rely on the detection of muscle activity, enabling users to control interfaces through intentional muscle contractions. Moreover, a limitation of EMG systems is the challenge of predicting muscle synergies \cite{Rugy}. While synergy-decomposition can reduce the dimensionality of EMG data, its ability to accurately predict muscle forces deteriorates as task complexity increases, which can further impact control performance \cite{Rugy}. The lack of adaptability in these single-modality BMIs forces users to rely on a fixed and limited control method, regardless of task complexity, reducing both usability and long-term adherence \cite{Vidal}. As a result, 30\% of users discontinue the use of assistive BMI devices within one year \cite{Phillips}.

Additionally, traditional BMIs that involve intricate control executions often also require complex decoders, which require extensive user training to generate data for training and calibrating the decoders. 

To overcome these limitations, this study presents a novel hybrid BMI that integrates SSVEP-based EEG signals and facial EMG to enhance control versatility, increase the degrees of freedom, and reduce user fatigue. By alternating between EEG and EMG modalities, the system allows users to rest muscle groups during EEG control and reduce cognitive load during EMG-based tasks, optimizing interaction efficiency while minimizing fatigue. To validate this approach, we implement a virtual interface where users navigate simulated turtles to designated areas on a screen, demonstrating real-time control using both neural and muscular signals.

This hybrid BMI offers greater task adaptability than existing single-modality systems by allowing users to leverage the strengths of each input method depending on the task demands. By dynamically switching between EEG and EMG control, this system enhances usability, reduces cognitive and physical strain, and provides a more practical, fatigue-resistant solution for individuals with severe motor impairments. Future work includes real-world validation with motor-impaired individuals and potential collaborations with assistive technology developers to translate this research into commercially viable solutions. This study aims to advance the field of BMI engineering by bridging the gap between flexibility, fatigue management, and task-specific adaptability, ensuring that assistive technology can better meet the needs of its users.

\section{Methods}

\begin{figure*}
  \centering
  \includegraphics[width=0.70\textwidth]{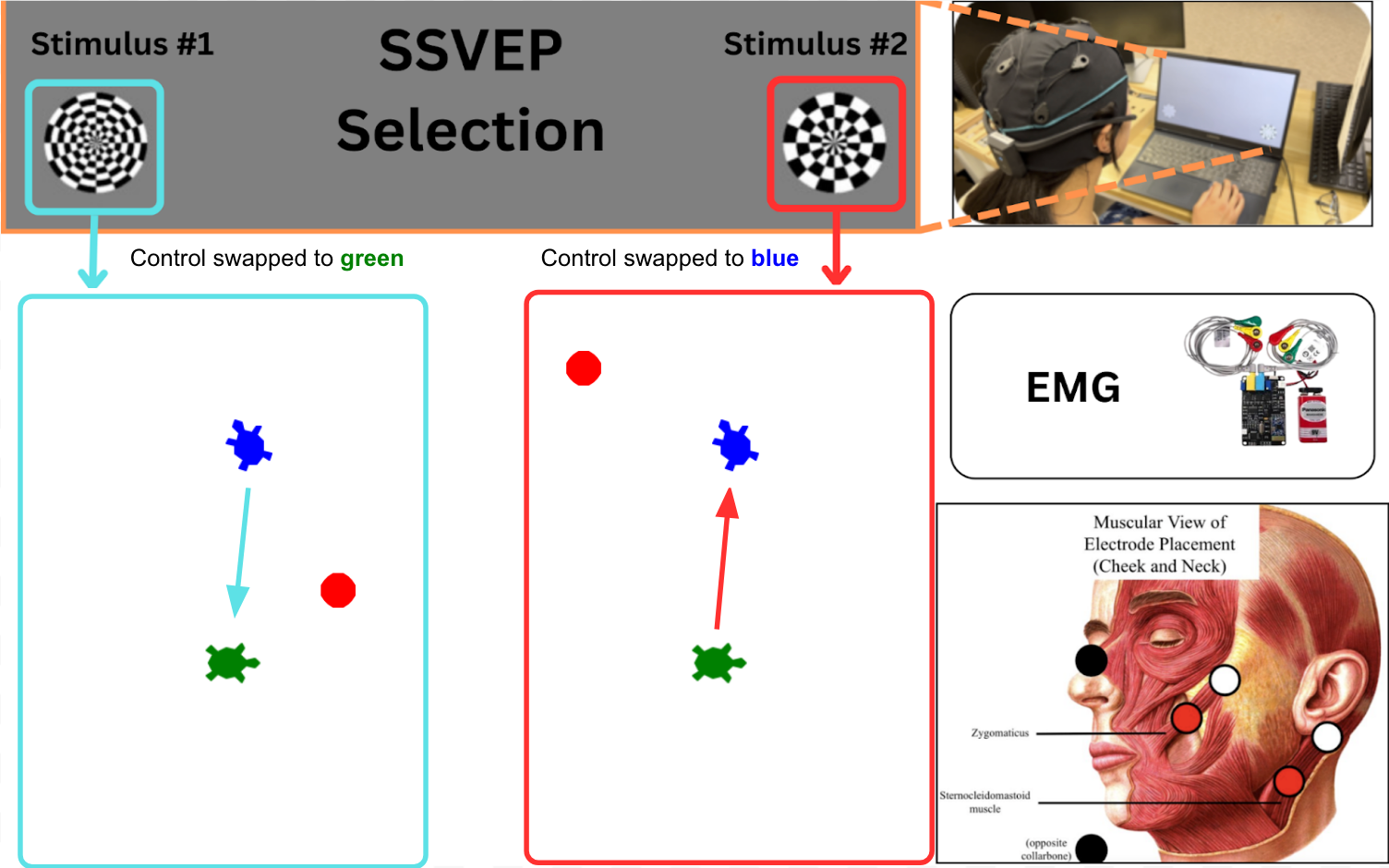}
  \caption{Staring at the left stimulus flashing at 9.75 Hz selected the blue turtle and staring at the right stimulus flashing at 14.25 Hz selected the green turtle. The EMG sensors corresponding to the colored circles were attached to the cheek muscle (zygomaticus) and the neck muscle (sternocleidomastoid) on opposite sides of the subject (as opposed to the same side that is displayed above) enabling control of the selected turtle.}
\label{fig:multimodal_experiment}
  
\end{figure*}

\subsection{Participants}
Twelve able-bodied human participants were recruited to test our hybrid system. All participants were between the ages of 18-21. Nine out of the twelve participants were first-time users of the system, while the other three subjects had used the system before the experiments. As this was our first round of testing, we chose to use able-bodied humans as our subjects so that we could validate the ease of use of the system before testing it out on our target demographic of individuals with motor disabilities. 

EMG signals were captured as electrical activity from contractions of the zygomaticus (hereafter referenced as the “cheek muscle”) and the sternocleidomastoid (hereafter referenced as the “neck muscle”) were recorded using surface electrodes. The signals were measured and streamed in real-time using Arduino Grove EMG sensors at a sampling frequency of 500 Hz and preprocessed using a 50-150 Hz band-pass filter to remove noise and improve signal quality.

\FloatBarrier
\subsection{Electromyography}
\begin{figure}[hbt!]
    \centering
    \includegraphics[width=0.475\textwidth]{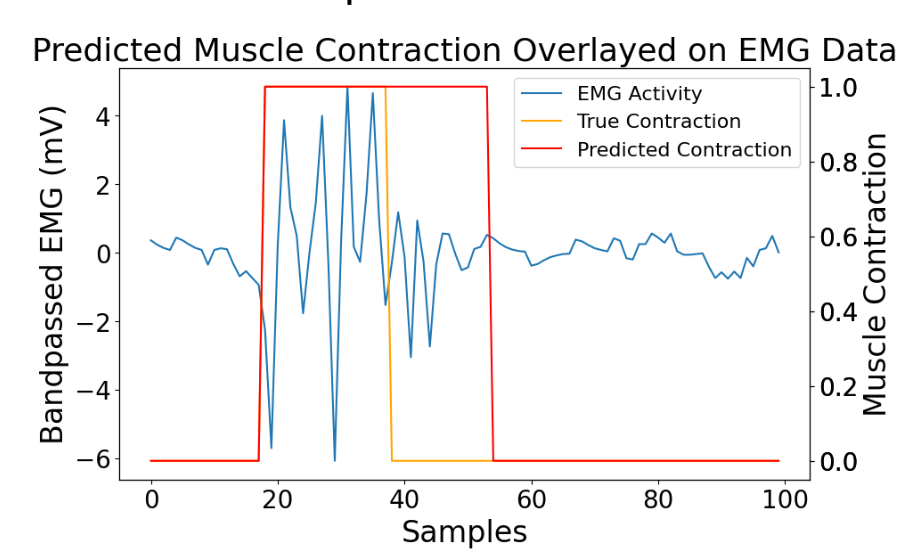}
    \caption{Predicted muscle contraction overlaid on bandpassed EMG data. The raw EMG activity (blue), true contraction that the subject labeled (yellow), and predicted contraction by the logistic regression model (red) are shown.}
    \label{fig:emg_decoding}
\end{figure}

While recording EMG electrode potential, we process each signal from the cheek and neck and classify them separately using pre-trained logistic regression models (Fig. \ref{fig:emg_decoding}). These regression models used a rolling window of 200ms of with an 80ms overlap between classification windows. When running the EMG decoding live we see a 95\% decoding accuracy when running live EMG decoding with pre-trained models. To pre-train EMG models, we collected EMG data by attaching a set of Grove EMG electrodes to both the neck and cheek muscles. The Grove EMG electrodes and a breadboard with two buttons attached are connected to an Arduino. Utilizing Teleplot and an Arduino script to record and bandpass filter EMG data, the participant's EMG data was simultaneously visualized and recorded for a duration of five minutes. Before the five minute recording period, there is a 10 second calibration period in which the subject does not move to establish a baseline. During the five minutes, participants are instructed to randomly flex a muscle while ensuring they leave a gap in between each muscle flex. While the participant was flexing a muscle, they were also instructed to hold down the button corresponding to the muscle they were flexing. 

\subsection{Electroencephalography}
\FloatBarrier
\begin{figure}[hbt!]
    \centering
    \includegraphics[width=0.475\textwidth]{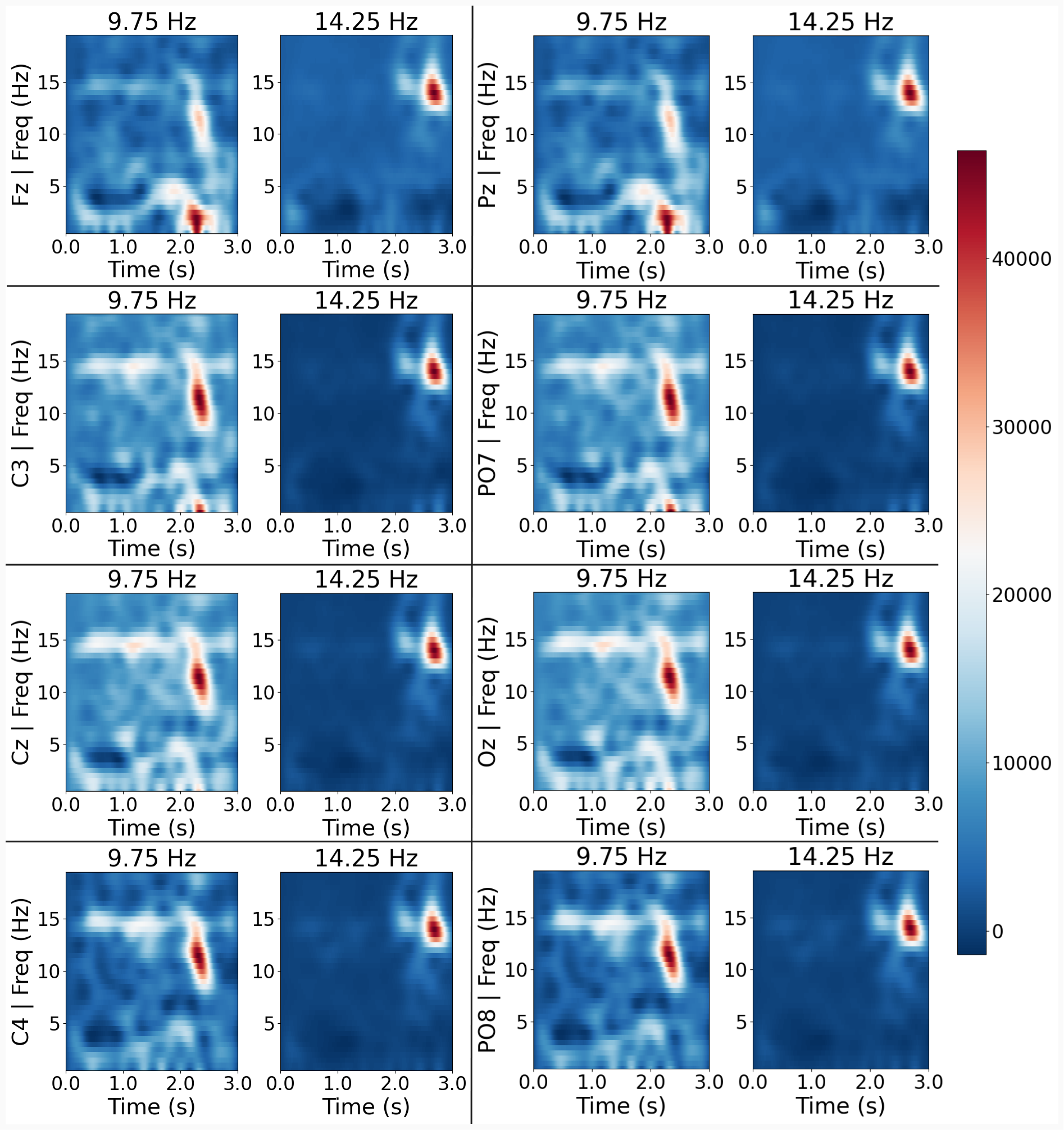}
    \caption{Time-frequency representation (TFR) of SSVEP data for subject 01. The TFR heat maps are generated for a window of 3 seconds for each of the 8 EEG channels when the subject is focusing on the 9.75 Hz and 14.25 Hz SSVEP stimulus respectively.}
    \label{fig:eeg_decoding}
\end{figure}

EEG data was measured and streamed via a g.tec Unicorn 8-channel wearable headset with a sampling rate of 250 Hz. The headset was utilized with wet electrodes. In the experimental task, the user is presented with two steady-state visually evoked stimuli (SSVEP) at 9.75 Hz and 14.25 Hz (Fig. \ref{fig:multimodal_experiment}). 

The EEG component was processed with a 2 Hz high pass filter and a 60 Hz notch filter using infinite impulse response (IIR) to preserve harmonics on a three second-time window. The processed EEG sequences are then classified using Canonical Correlation Analysis (CCA) to detect SSVEP signals at 9.75 and 14.25 Hz that were emitted in the visual cortex with a three second time window. We measured the CCA correlations associated with the 9.75 Hz, $C_{9.75}$, and the 14.25 Hz, $C_{14.25}$, signal (Fig. \ref{fig:eeg_decoding}).

\begin{figure}[hbt!]
    \centering
    \includegraphics[width=0.475\textwidth]{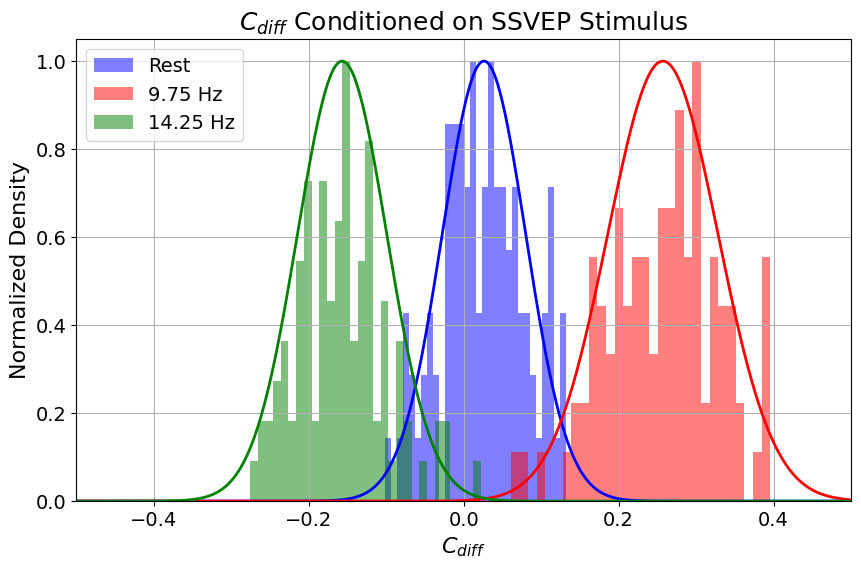}
    \caption{Histogram and Gaussian fit of $C_{\text{diff}}$ under different SSVEP stimuli: rest (blue), 9.75 Hz (red), and 14.25 Hz (green). Solid lines show Gaussian fits, with the x-axis representing $\mathrm{C_{\text{diff}}}$ and the y-axis showing normalized density.}
    \label{fig:cond_dist_ssvep}
\end{figure}

A probability distribution was constructed on the differences between the CCA correlation of the 9.75 vs 14.25 Hz stimulus $C_{\text{diff}} = C_{9.75} - C_{14.25}$ conditioned on the subject attention, or lack thereof, to a particular visual stimulus. This was done to classify when the subject was in a “rest” state, ie when the subject was not attending to either stimulus. To calibrate the conditional distribution 50 iterations of $C_{\text{diff}}$ were recorded when the subject fixated on a blank screen without stimulus, on a screen with only the 9.75 Hz stimulus was present, and on a screen with only the 14.25 Hz stimulus was present, respectively. A Gaussian approximation of the recorded conditional distributions was then constructed (Fig. \ref{fig:cond_dist_ssvep}).
Since precision is more highly valued in a user experience, in order to prevent false positives, a threshold of 2 standard deviations away from the mean of the rest state was implemented to classify each stimulus.

\subsection{Experimental Design}

The Python Turtle standard library was used to create a graphical interface in which the user controlled two turtle-shaped icons that each had two dimensions of freedom (Fig. \ref{fig:multimodal_experiment}). The goal of the task was to move one of the graphical turtles to a red dot (called a ‘ball’) on the screen. The two turtles start in the middle of the screen. The blue one starts toward the left and the green one starts towards the right. After every successful effort, the ball was then moved to another corner of the screen in a clockwise fashion. A trial involved moving a turtle to the ball 5 times, starting at the top left corner. 

The user had three available actions in this task. Contraction of the cheek would toggle on and off forward linear motion. Contraction of neck muscles would toggle on and off rotational motion upon the turtle’s central axis. The third action was triggered by sequential contraction of the cheek muscle followed by the neck muscle, and resulted in the switching of control of one turtle to the other. Before data collection, users were allowed one trial to understand the rules. During data collection, 5 trials of collecting 5 balls were completed. 

In the hybrid EMG and EEG control paradigm, the linear and rotational movements of individual turtles remained consistent as the EMG-only paradigm. However, a second display was provided to the user that showed the SSVEP stimuli as discussed previously. Fixation on the 9.75 Hz stimulus gave the user the ability to switch to a blue turtle, while fixation on the 14.25 Hz stimulus gave them control over the green one. Upon recognition of an SSVEP signal by the EEG decoding pipeline, an auditory stimulus playing “GREEN” or “BLUE” respectively, indicated to the user which turtle they now had control over.

\section{Results and Discussion}

\FloatBarrier

\begin{figure}[hbt!]
    \centering
    \includegraphics[width=0.475\textwidth]{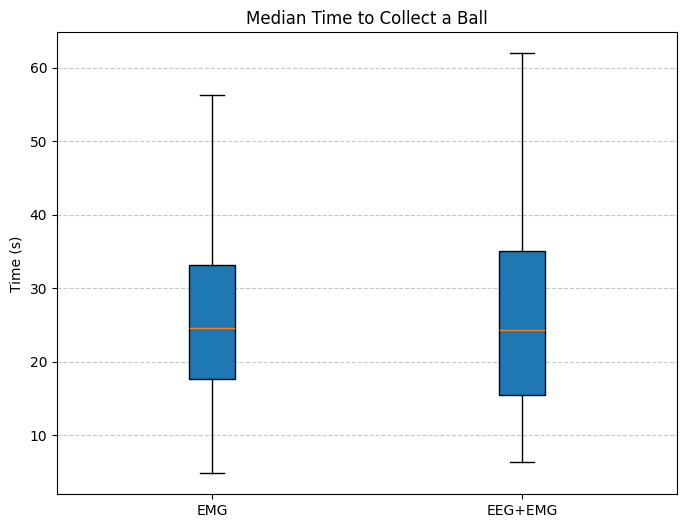}
    \caption{Box and whisker plot showing median time and interquartile range to collect a single ball across different subjects using the EMG-only and hybrid control paradigm respectively.}
    \label{fig:experimental_results}
\end{figure}

The time required for each subject to collect a ball was recorded across all experimental conditions. The study involved twelve participants, each completing 5 trials per condition, with each trial requiring them to control the virtual turtles to collect 5 balls.

To establish a baseline, we measured performance in the EMG-only condition, where subjects achieved a median completion time of 24.67 seconds. When using the hybrid EMG and EEG system, subjects maintained similar performance levels, achieving a median completion time of 24.29 seconds (Fig. \ref{fig:experimental_results}). These results demonstrate that integrating EEG alongside EMG does not negatively impact task performance, supporting the feasibility of multimodal control without introducing additional cognitive or physical strain on users.

User fatigue was assessed using the NASA Task Load Index (NASA-TLX) \cite{HartNasa}, which measures metrics such as physical demand and effort. Results from the NASA-TLX showed that about 90\% of subjects indicated that using the hybrid system matched or improved the physical demand and effort needed to control the machine effector compared to the EMG only control paradigm.

This study introduces a general-purpose hybrid EEG-EMG control framework capable of managing systems with four degrees of freedom. The turtle-based interface served as a proof-of-concept to validate the usability of this multimodal paradigm in first-time users. This framework can be extended to various machine effectors requiring independent control over multiple movement axes. For instance, in a drone control system, one SSVEP stimulus could toggle vertical motion, with EMG signals directing ascent and descent, while a second SSVEP stimulus could toggle planar motion, with EMG signals dictating linear and rotational movement. This adaptable control structure highlights the system’s potential for a wide range of robotic applications.

Beyond its engineering implications, this hybrid BMI framework has significant potential in clinical assistive technologies. For individuals with ALS, spinal cord injuries, or other neuromuscular disorders, the ability to interact with robotic devices in a fatigue-minimized manner could enhance daily autonomy. Potential applications include wheelchair navigation, robotic prosthesis control, and telepresence robotics, allowing users with severe motor impairments to engage with their environments more effectively.

A key advantage of this hybrid approach is its ability to distribute control efforts across multiple physiological channels, thereby reducing strain on any single input source. By alternating between EEG and EMG, the system mitigates cognitive and physical fatigue, making it more practical for prolonged use. These findings validate the viability of multimodal BMI control and pave the way for future implementations in high-dimensional control tasks, expanding the scope of assistive technology and neuroprosthetic applications.

\section{Conclusions}

This study presents a novel hybrid BMI system that seamlessly integrates SSVEP-based EEG and facial EMG, enhancing control versatility and minimizing fatigue in assistive technologies. By dynamically alternating between neural and muscular signals, the system optimizes interaction efficiency while reducing both cognitive and physical strain. Experimental results confirm that the hybrid BMI maintains comparable task completion times to an EMG-only system while improving user-reported physical demand and effort, demonstrating its potential for prolonged usability.

These findings underscore the advantages of multimodal BMIs in improving usability, adaptability, and long-term adherence, addressing the inherent limitations of single-modality systems. This work establishes a foundation for further exploration into real-world applications, particularly for individuals with severe motor impairments. Future research will focus on validating the system with motor-impaired users, refining signal processing techniques for enhanced accuracy, and expanding its applicability to complex control tasks, such as drone piloting and robotic systems. 

By bridging the gap between flexibility, fatigue management, and task-specific adaptability, this study advances BMI engineering and contributes to the development of more user-friendly and effective assistive devices, paving the way for broader adoption in neuroprosthetics, mobility aids, and teleoperated systems.

\section*{Acknowledgment}

We thank the JHU Brain-Computer Interface Society members for their dedication and contributions. We are also sincerely grateful for the generous financial support from the JHU HOUR Office and the JHU WSE Alumni Association.

\bibliographystyle{IEEEtran}
\bibliography{References}

\begin{thebibliography}{10}
\providecommand{\url}[1]{#1}
\csname url@samestyle\endcsname
\providecommand{\newblock}{\relax}
\providecommand{\bibinfo}[2]{#2}
\providecommand{\BIBentrySTDinterwordspacing}{\spaceskip=0pt\relax}
\providecommand{\BIBentryALTinterwordstretchfactor}{4}
\providecommand{\BIBentryALTinterwordspacing}{\spaceskip=\fontdimen2\font plus
\BIBentryALTinterwordstretchfactor\fontdimen3\font minus \fontdimen4\font\relax}
\providecommand{\BIBforeignlanguage}[2]{{%
\expandafter\ifx\csname l@#1\endcsname\relax
\typeout{** WARNING: IEEEtran.bst: No hyphenation pattern has been}%
\typeout{** loaded for the language `#1'. Using the pattern for}%
\typeout{** the default language instead.}%
\else
\language=\csname l@#1\endcsname
\fi
#2}}
\providecommand{\BIBdecl}{\relax}
\BIBdecl

\bibitem{Shih}
\BIBentryALTinterwordspacing
J.~J. Shih, D.~J. Krusienski, and J.~R. Wolpaw, ``Brain-computer interfaces in medicine.'' \emph{Mayo Clinic proceedings}, vol. 87 3, pp. 268--79, 2012. [Online]. Available: \url{https://api.semanticscholar.org/CorpusID:10447364}
\BIBentrySTDinterwordspacing

\bibitem{Farina}
\BIBentryALTinterwordspacing
D.~Farina, N.~Jiang, H.~Rehbaum, A.~Holobar, B.~Graimann, H.~Dietl, and O.~C. Aszmann, ``The extraction of neural information from the surface emg for the control of upper-limb prostheses: Emerging avenues and challenges,'' \emph{IEEE Transactions on Neural Systems and Rehabilitation Engineering}, vol.~22, pp. 797--809, 2014. [Online]. Available: \url{https://api.semanticscholar.org/CorpusID:22300472}
\BIBentrySTDinterwordspacing

\bibitem{Nicols}
\BIBentryALTinterwordspacing
L.~F. Nicol{\'a}s-Alonso and J.~G. Gil, ``Brain computer interfaces, a review,'' \emph{Sensors (Basel, Switzerland)}, vol.~12, pp. 1211 -- 1279, 2012. [Online]. Available: \url{https://api.semanticscholar.org/CorpusID:12187240}
\BIBentrySTDinterwordspacing

\bibitem{Young}
\BIBentryALTinterwordspacing
A.~J. Young, L.~J. Hargrove, and T.~A. Kuiken, ``The effects of electrode size and orientation on the sensitivity of myoelectric pattern recognition systems to electrode shift,'' \emph{IEEE Transactions on Biomedical Engineering}, vol.~58, pp. 2537--2544, 2011. [Online]. Available: \url{https://api.semanticscholar.org/CorpusID:26408204}
\BIBentrySTDinterwordspacing

\bibitem{Trejo}
\BIBentryALTinterwordspacing
L.~J. Trejo, R.~Kochavia, K.~Kubitzb, L.~D. Montgomerya, R.~Rosipala, and B.~Matthewsa, ``Eeg-based estimation of cognitive fatigue,'' \emph{Psychology}, vol.~6, 2015. [Online]. Available: \url{http://dx.doi.org/10.4236/psych.2015.65055}
\BIBentrySTDinterwordspacing

\bibitem{Cao}
\BIBentryALTinterwordspacing
T.~Cao, F.~Wan, C.~M. Wong, J.~N. da~Cruz, and Y.~Hu, ``Objective evaluation of fatigue by eeg spectral analysis in steady-state visual evoked potential-based brain-computer interfaces,'' \emph{BioMedical Engineering OnLine}, vol.~13, pp. 28 -- 28, 2014. [Online]. Available: \url{https://api.semanticscholar.org/CorpusID:2332770}
\BIBentrySTDinterwordspacing

\bibitem{Rugy}
\BIBentryALTinterwordspacing
A.~de~Rugy, G.~E. Loeb, and T.~J. Carroll, ``Are muscle synergies useful for neural control?'' \emph{Frontiers in Computational Neuroscience}, vol.~7, 2013. [Online]. Available: \url{https://api.semanticscholar.org/CorpusID:868277}
\BIBentrySTDinterwordspacing

\bibitem{Vidal}
\BIBentryALTinterwordspacing
G.~Vidal, M.~L. Rynes, Z.~Kelliher, and S.~J. Goodwin, ``Review of brain-machine interfaces used in neural prosthetics with new perspective on somatosensory feedback through method of signal breakdown,'' \emph{Scientifica}, vol. 2016, 2016. [Online]. Available: \url{https://api.semanticscholar.org/CorpusID:14877773}
\BIBentrySTDinterwordspacing

\bibitem{Phillips}
\BIBentryALTinterwordspacing
B.~T. Phillips and H.~Zhao, ``Predictors of assistive technology abandonment.'' \emph{Assistive technology : the official journal of RESNA}, vol. 5 1, pp. 36--45, 1993. [Online]. Available: \url{https://api.semanticscholar.org/CorpusID:27759983}
\BIBentrySTDinterwordspacing

\bibitem{HartNasa}
\BIBentryALTinterwordspacing
S.~G. Hart, ``Nasa-task load index (nasa-tlx); 20 years later,'' \emph{Proceedings of the Human Factors and Ergonomics Society Annual Meeting}, vol.~50, pp. 904 -- 908, 2006. [Online]. Available: \url{https://api.semanticscholar.org/CorpusID:6292200}
\BIBentrySTDinterwordspacing

\end{thebibliography}

\end{document}